
\magnification=\magstep1
\font\title=cmbx10

\font\name=cmr8

\hsize=6.5truein
\vsize=9.5truein
\baselineskip=6.5mm
\parindent=6.5mm
\hskip 10.5cm
\null
{\nopagenumbers
\smallskip\noindent
\hskip 10.5cm LMU-TPW 92-8\vskip .5pt\noindent
\hskip 10.5cm May 1992
\bigskip\noindent
\centerline{\title FREE AND INTERACTING 2-D MAXWELL FIELD THEORY}
\centerline{\title ON CONFORMALLY FLAT SPACE-TIMES}
\vskip 2cm
\centerline{F. F{\name ERRARI}\footnote{\name ${}^a$}{\name Work supported
by the Consiglio Nazionale Ricerche, P.le A. Moro 7, Roma, Italy}}\smallskip
\centerline{\it Sektion Physik der Universit\"at M\"unchen}\smallskip
\centerline{\it Theresienstr. 37, 8000 M\"unchen 2}\smallskip
\centerline{\it Fed. Rep. Germany}
\vskip 3truecm
\centerline{\title Abstract}
\vskip 1truecm
{\narrower The free Maxwell field theory is quantized in the Lorentz gauge
on a two dimensional manifold $M$ with conformally flat background metric.
It is shown that in this gauge the theory is equivalent, at least
at the classical level, to a
biharmonic version
of the bosonic string theory.
This equivalence is exploited in order to construct in details the propagator
of the Maxwell field theory on $M$.
The expectation values of the Wilson loops are computed.
A trivial result is obtained confirming in the Lorentz gauge previous
calculations.
Finally the interacting case is briefly discussed taking the Schwinger model
as an example. The two and three point
functions of the Schwinger model are explicitly derived at the lowest order
on a Riemann surface.}
\vfill\eject}
\pageno=1
\centerline{\title 1. INTRODUCTION}
\vskip 1cm
In recent times the two dimensional gauge invariant field
theories have been a source of many interesting developments [1].
This paper concerns
the quantization of the Maxwell Field Theory (MFT) on a two dimensional
Riemannian manifold admitting conformally flat background metrics.
Classically the MFT, like the more complicated Yang-Mills theories defined
on Riemann surfaces, is well understood [2]. The quantum case however is
not so well known. Earlier results about the Yang-Mills theories
on the sphere and on the
cylinder are given in [3].
On a general Riemann surface the MFT has been quantized for example
in [4,5], where however the equations of motion were used
in order to compute the vacuum expectation values (VEV's)
of the gauge invariant quantities like the Wilson loops (WL).
\smallskip
Here we perform a full quantization of the MFT on a general two dimensional
complex manifold. As examples the complex plane, the disk and the
closed and orientable Riemann surfaces of any genus are considered.
We have found very convenient
to quantize the theory in the Lorentz gauge.
In this way, in fact, the transversal and longitudinal components of the
fields are decoupled (see e.g. [6]) and the transversal part can be
expressed in terms of a purely imaginary scalar field.
\smallskip
The material contained in this paper is organized as follows:
\vskip 1pt\noindent
In Section 2 the MFT on a two dimensional manifold is introduced using a set
of complex coordinates.
After choosing a conformally flat background metric, the Lorentz gauge is
imposed so that just the trasverse field can propagate.
Exploiting the Hodge decomposition of a general one form
in exact, coexact and harmonic forms, it is shown that the MFT in the Lorentz
gauge is equivalent to a theory of scalar fields $\varphi(z,\bar z)$
with higher order derivatives.
The residual gauge invariance, characteristic of the Lorentz gauge fixing, is
analyzed and the proper boundary conditions in order to get rid of it
are assigned.\smallskip
In Section 3 the flat case is investigated. The biharmonic equations of motion
for the scalar fields $\varphi$ are solved on a disk and on the complex
plane and the explicit form of the propagators is derived.
The difficulties of defining the propagators on noncompact manifolds are
pointed out [7].\smallskip
In Section 4 the results of Section 3 are extended to the case of the Riemann
surfaces. The VEV of a WL is computed giving the trivial result
of [4,5].\smallskip
In Section 5 the MFT interacting with a theory of massless
fermions is studied in the path integral formalism.
The explicit calculations of the physical amplitudes are simplified
by the fact that the longitudinal gauge fields can be integrated away.
The result of this integration is a $\delta$-functional expressing
the physical requirement of the conservation of the fermionic number
in the amplitudes.
The computation of the two and three point functions of the Schwinger
model on a Riemann surface is performed at the lowest order of the
perturbation theory.\smallskip
Finally Section 6 is a discussion about the possibility of coupling the MFT
to the $b-c$ systems and to the massless scalar fields of string theory.
A model with higher order derivatives and with nonAbelian gauge group
of symmetry is introduced
on the complex sphere.
\vskip 1cm
\centerline{\title 2. MAXWELL FIELD THEORY ON A COMPLEX BACKGROUND}
\vskip 1cm
On a general two-dimensional complex manifold $M$ we consider
the following functional:
$$S[J_\mu,A_\mu]={1\over 4}\int_{M}d^2x\sqrt{-g}\left (
F_{\mu\nu}F^{\mu\nu} +J_\mu A^{\mu}\right)\eqno(2.1)$$
where $F_{\mu\nu}=\partial_\mu A_\nu-\partial_\nu A_\mu$,
$\mu,\nu=1,2$, is the usual field strength and $g_{\mu\nu}$ is a background
metric with Minkowski signature.
Finally $J_\mu(x)$ represents an external current.
As an explicit example of two-dimensional manifold $M$
we consider the closed and orientable Riemann surface
of genus $g$, $\Sigma_g$.
On $\Sigma_g$ we define a canonical set of independent homology cycles
$A_i$, $B_j$, $i,j=1,\ldots,g$.
The classical
equations of motion of the fields $A_\mu$ are:
$$(\sqrt{-g})^{-1}\partial_\nu\left (\sqrt{-g} F_{\mu\nu}(x)\right )=J^\mu(x)
\eqno(2.2)$$
If we put $J^\mu(x)$=0 and consider the antisymmetric character of the field
strength $F^{\mu\nu}(x)$, eq. (2.2) implies (see e.g. [5]):
$$F={1\over 2}[\epsilon]_{\mu\nu}F^{\mu\nu}=\alpha\eqno(2.3)$$
where $\alpha$ is an arbitrary constant and $[\epsilon]_{\mu\nu}$
is the Levi-Civita tensor.
Due to the fact that $F^{\mu\nu}$ has only one nonvanishing component in
two dimensions, namely $F^{12}$, eq. (2.3) implies that the MFT described by
eq. (2.1) has just discrete degrees of freedom depending on the topology of
the background.
In [4] eq. (2.3) was imposed also in the quantized version of the MFT inserting
by hand the constraint $F-\alpha=0$ in the path integral.
Also in [5] eq. (2.3) was used in the computation of the VEV of the Wilson
loops. In this case the classical constraint (2.3) was exploited in
computing the
commutator of the anti-BRST operator with the ghost fields.
However, eq. (2.3) does not hold in general when the fields are quantum
operators even when the two dimensional space-time has the topology
of ${\bf R}^2$.\smallskip
In order to keep our results as general as possible we will forget in the
following eq. (2.3).
Before to quantize the MFT, we perform a Wick rotation in the $x^2$ axis:
$x^2\rightarrow ix^2$ and we choose a system of local complex coordinates on
$M$:
$$\cases{z=x^1+ix^2\cr {\bar z}=x^1-ix^2\cr}\quad\quad
\cases{\partial_z={1\over 2}(\partial_1-i\partial_2)\cr
\partial_{\bar z}={1\over 2}(\partial_1+i\partial_2)\cr}\eqno(2.4)$$
In this paper we will express the fields and their functionals
using a local system of coordinates like that of eq. (4).
The final results, as the propagators and the solutions of the
equations of motion, will be however globally defined.\smallskip
The nonvanishing component of the field strength $F_{\mu\nu}$ is:
$$F_{z{\bar z}}=\partial_z A_{\bar z}-\partial_{\bar z}A_z\eqno(2.5)$$
Clearly the complex field strength
${1\over 2}F_{z{\bar z}}dz\wedge \bar z$
does not need the introduction of the covariant
derivatives $\nabla_z$ and $\nabla_{\bar z}$ in order to transform
covariantly under general diffeomorphisms.
In complex coordinates the action (2.1) becomes:
$$S[A_z,A_{\bar z};J_z,J_{\bar z}]={i\over 4}\int_M\sqrt{g}\left(
F_{z{\bar z}}F^{z{\bar z}}+J_zA^z+J_{\bar z}A^{\bar z}\right)$$
In this equation the factor $i$ coming from the Wick rotation has been
neglected.
We will restore it below as the path  integral quantization will be
introduced.
One can simplify the above equation as follows:
$$S[A_z,A_{\bar z},J_z,J_{\bar z}]={i\over 4}\int_M\left[
{1\over \sqrt{g}}F_{z\bar z}^2+\sqrt{g}J_zA^z+\sqrt{g}J_{\bar z}A^{\bar z}
\right ]\eqno(2.6)$$
At this point we choose a conformally flat metric:
$$g_{zz}=g_{{\bar z}{\bar z}}=0\qquad\qquad g_{z{\bar z}}={1\over 2}
e^{2\sigma(z,{\bar z})}\eqno(2.7)$$
where $\sigma(z,{\bar z})$ is a real function of $z$ and ${\bar z}$.
The only effect in eq. (2.6) is that now $\sqrt{g}=g_{z{\bar z}}$
on a local patch.
Substituting eqs. (2.5) and (2.7) in (2.6) we get:
$$S[A,J]={1\over 2}\int_Md^2zg^{z{\bar z}}\left (\partial_zA_{\bar z}
\partial_{\bar z}A_z-\partial_zA_{\bar z}\partial_zA_{\bar z}+
g^{z\bar z}J_zA_{\bar z}
+{\rm c.c.}\right )\eqno(2.8)$$
Note that the external current $J_\mu$ has been rescaled by a constant
factor $1/2$.
{}From eq. (2.8) the equations of motion ${\delta S\over\delta A_z}=
{\delta S\over\delta A_{\bar z}}=0$ for the fields $A_z$ and
$A_{\bar z}$ become:
$$\cases{\partial_z\left (-g^{z{\bar z}}\partial_{\bar z}A_z+
g^{z{\bar z}}\partial_zA_{\bar z}\right)=J_z\cr
\partial_{\bar z}\left(-g^{z{\bar z}}\partial_zA_{\bar z}+g^{z{\bar z}}
\partial_{\bar z}A_z\right)=J_{\bar z}\cr}\eqno(2.9)$$
Now it is possible to quantize the MFT using the path integral
formalism:
$$Z[J]=\int DA_zDA_{\bar z}e^{-S[A,J]}\eqno(2.10)$$
where $S[A,J]$ is the action of eq. (2.8).\smallskip
The above partition function is not well defined due to the $U(1)$ gauge group
of symmetry:
$$\eqalign{A_z\rightarrow & A_z+\partial_z\lambda(z,{\bar z})\cr
A_{\bar z}\rightarrow & A_{\bar z}+\partial_{\bar z}\lambda(z,{\bar z})\cr}
\eqno(2.11)$$
where $\lambda(z,{\bar z})$ is a real function of $z$ and ${\bar z}$.
In order to get rid of the spurious integration over the gauge degrees of
freedom we impose the Lorentz gauge. To this purpose we express the vector
fields $A_z$ and $A_{\bar z}$ in terms of a complex scalar field
$\chi(z,\bar z)$ as follows:
$$A_z(z,{\bar z})={1\over 2}\partial_z\chi(z,{\bar z})+A_z^{\rm har}
\qquad\qquad
A_{\bar z}={1\over 2}\partial_{\bar z}{\bar \chi}(z,{\bar z})
+A_{\bar z}^{\rm har}\eqno(2.12)$$
In eq. (2.12) $A_\mu^{\rm har}$, $\mu=z,\bar z$, denotes the zero mode content
of a general vector field $A_\mu$. On a Riemann surface we have [8]:
$$A_z^{\rm har}=\sum\limits_{i,k=1}^g2\pi i\bar u_k(\Omega-\bar
\Omega)^{-1}_{kj}(\omega_z)_j(z)$$
where $\Omega$ is the period matrix, $\bar u_k$ is a vector with $g$
real components and $(\omega_z)_i(z)$, $i=1,\ldots,g$, is a basis of zero
modes normalized as follows:
$$\oint_{A_i}\omega_jdz=0\qquad\qquad \Omega_{ij}=\oint_{B_i}\omega_j(z)dz$$
Let us denote with $\Re[T]$ and $\Im[T]$ the real and imaginary parts of a
tensor $T$ respectively.
If we put
$$\rho=\Re[\chi]\qquad\qquad\qquad \varphi=i\Im[\chi]\eqno(2.13)$$
so that $\chi=\varphi+\rho$, eq. (2.12) turns out to be the Hodge
decomposition of a general vector field in the Euclidean space:
$$\eqalign{
A_z=&{1\over 2}\partial_z\varphi+{1\over 2}\partial_z\rho+A_z^{\rm har}\cr
A_{\bar z}=&-{1\over 2}\partial_{\bar z}+{1\over 2}\partial_{\bar z}\rho
+A_{\bar z}^{\rm har}\cr}\eqno(2.14)$$
The advantage of the decomposition (2.14) in our case is that the vectors
$A_z^T=\partial_z\varphi$ and $A_z^L=\partial_z\rho$ describe, with their
complex conjugate partners, the transverse and longitudinal components
of $A_\mu$ respectively.
This can be easily seen in the Minkowski space, where eq. (2.14)
becomes:
$$A_\mu={1\over 2}\epsilon_{\mu\nu}\partial^{\nu}\varphi+{1\over 2}
\partial_\mu\rho+A_\mu^{\rm har}$$
In the Euclidean space it is possible to check that the free part of the
action (2.8), i.e. with $J_\mu$ set to zero, does not depend
on the longitudinal vector fields $A_z^L$ and $A_{\bar z}^L$.
One of the important properties of the decomposition (2.14) is that,
in the scalar product $(A_\mu,A_\nu)\equiv\int d^2zA_zA_{\bar z}$, the
three different components of $A_z$, namely $A^T_z$, $A_z^L$ and
$A_z^{\rm har}$ are mutually orthogonal.
This orthogonality property will be extensively used in the following.
Now we are ready to exploit the substitution (2.14) in the path integral
(2.10). As in the case of flat space [6], it turns out that the integration
over the unphysical longitudinal fields $A_z^L$, $A_{\bar z}^L$ can be easily
performed leaving us only with the transverse fields $A_z^T$, $A_{\bar z}^T$.
We notice that this way of fixing the gauge is equivalent of having chosen
the Lorentz gauge fixing.
As a consequence it remains a residual gauge invariance proper of the Lorentz
gauge that will be discussed later.
\smallskip
In order to proceed, we have to
decompose also the external currents as we did for the
vector fields in eq. (2.14):
$$\cases{J_z^T=\partial_zJ\cr
J_{\bar z}^T=-\partial_{\bar z}J\cr}\qquad\qquad
\cases{J_z^L=\partial_z{\tilde J}\cr
J_{\bar z}^L=\partial_{\bar z}{\tilde J}\cr}\eqno(2.15)$$
where $J$ is a purely imaginary scalar field and ${\tilde J}$ is purely real.
At this point we substitute eqs. (2.14)-(2.15) in eq. (2.10).
The functional measure in the path integral becomes:
$$DA_zDA_{\bar z}=D\varphi D\rho\enskip {\rm det}\left|
\eqalign{{\delta A_z\over \delta\varphi}&{\delta A_{\bar z}\over\delta\varphi}
\cr
{\delta A_z\over \delta\rho}&{\delta A_{\bar z}\over\delta\rho}\cr}\right|=
D\varphi D\rho \enskip {\rm det}(2
\partial_z\partial_{\bar z})\eqno(2.16)$$
After a simple calculation eq. (2.10) yields:
$$Z[J,{\tilde J}]=\int D\varphi D\rho{\rm det}(
\partial_z\partial_{\bar z})e^{\int_M
{1\over 4}d^2z\left[ g^{z{\bar z}}\partial_z\partial_{\bar z}\varphi
\partial_z\partial_{\bar z}\varphi+\partial_zJ\partial_{\bar z}\varphi
+\partial_z{\tilde J}\partial_{\bar z}\rho+A_z^{\rm har}J_{\bar z}^{\rm har}
+{\rm c.c.}\right]}$$
Notice that there is no dependence on the harmonic sector since we are free
to set $J_z^{\rm har}=J_{\bar z}^{\rm har}=0$.
The integrations over $D\rho$ and over the discrete degrees of freedom coming
from $A^{\rm har}$ is trivial and the result can be factored out together
with ${\rm det}(\partial_z\partial_{\bar z})$ which, in this context,
may be considered as a constant
depending on the moduli.
At the end we get:
$$Z[J]=\int D\varphi e^{S[\varphi,J]}\eqno(2.17)$$
where
$$S[\varphi,J,{\tilde J}]=\int_Md^2z\left[{g^{z\bar z}\over 4}
\triangle_z\varphi\triangle_z\varphi+\partial_zJ\partial_{
\bar z}\varphi+{\rm c.c.}\right]\eqno(2.18)$$
In the following we will use the convenient notation $\triangle_z=
\partial_z\partial_{\bar z}$.
The positive sign with which the action in eq. (2.17) is exponentiated
is correct if we remember that the field $\varphi$ is purely imaginary
as it follows from its definition in eq. (2.13).
Eq. (2.18) can be also expressed in terms of the transverse fields
$A^T_z(z,{\bar z})$.
After some integrations by parts in the sector with the external currents
we have:
$$Z[J_{\bar z}^T]=\int DA_z^T{\rm det}(\partial_{\bar z})
e^{\int_Md^2z{1\over 2}g^{z{\bar z}}\left[(\partial_{\bar z}A_z^T
\partial_{\bar z}A^T_z)+g_{z\bar z}J_{\bar z}^TA_z^T\right]}\eqno(2.19)$$
Remembering that
$\partial_{\bar z}A^T_z=-\partial_zA^T_{\bar z}$ due to eq. (2.14), we can
cast the action appearing in eq. (2.19) in the more usual form:
$$S[A_z^T,J_{\bar z}^T]=\int_Md^2z\left[-{g^{z\bar z}\over 2}
-\partial_{\bar z}A^T_z\partial_zA^T_{\bar z}+J_{\bar z}^T
A^T_z+{\rm c.c.}\right]$$
Apparently, there is just one degree of freedom in eq. (2.19) due to the fact
that $A_{\bar z}^T={\overline{A_z^T}}$.
However, as it happens in the case of Chern-Simons field theory
[9], one can show that also
the remaining degree of freedom disappears
in the canonical quantization.
The equations of motion of the fields $A^T_z$ and $\varphi(z,\bar z)$ are
$$\partial_{\bar z}g^{z\bar z}\partial_{\bar z}A_z^T=J_{\bar z}^T\eqno(2.20)$$
$$\triangle_zg^{z\bar z}\triangle_z\varphi=
\triangle_zJ\eqno(2.21)$$
Apart from the presence of the metric tensor sandwiched
between the two D'Alembertians, (2.21) is a biharmonic equation.
The presence of the metric is however unavoidable when one has to define the
biharmonic equation on a general manifold. A manifold is in fact covered by
many charts and without the metric the operator $\triangle_z^2$ is no longer
invariant after a tranformation of local coordinates.\smallskip\noindent
In the next section we will need the propagator of the scalar fields
$G(z,w)=<\varphi(z)\varphi(w)>$. $G(z,w)$ satisfies the following
equations:
$$\cases{\triangle_zg^{z\bar z}\triangle_z
G(z,w)=\delta_{z\bar z}^{(2)}(z,w)\cr
\triangle_wg^{w\bar w}\triangle_wG(z,w)=
\delta_{w\bar w}^{(2)}(z,w)\cr}\eqno(2.22)$$
In eq. (2.22) we have not ta\-ken into ac\-count the pos\-si\-ble zero
modes
of the op\-er\-a\-tor $\triangle_z g^{z\bar z}
\triangle_z$
which will be discussed later.
Since the field $\varphi(z,\bar z)$ is purely imaginary, we should in
principle put a minus
sign in front of the $\delta$-functions of eq. (2.22).
However our choice of sign is motivated by the fact that eventually we
are interested only in the propagator of the physical fields $A_z^T$,
$A_{\bar z}^T$.
Once we have the propagator of the scalar fields we can construct in fact also
the propagator of the vector fields $A_z^T$:
$$<A_z^T(z,\bar z)A_w^T(w,\bar w)>=\partial_z\partial_wG(z,w)\eqno(2.23)$$
It is easy to show that
$$\partial_{\bar z}g^{z\bar z}\partial_{\bar z}
<A_z^T(z,\bar z)A_w^T(w,\bar w)>=-\delta_{z\bar z}^{(2)}(z,w)$$
noting that $A_w^T(w,\bar w)=\partial_w\varphi(w,\bar w)$
and that the derivative
$\partial_w$ commutes with the metric $g^{z\bar z}$.
The desired minus sign in the above equation is produced remembering that
locally one has:
$$\partial_z\partial_{\bar z}G(z,w)=g_{z\bar z}{\rm log}|z-w|^2$$
and that $\partial_w\partial_{\bar z}{\rm log}|z-w|^2=-\delta^2_{z\bar z}
(z,w)$.
The form of the propagator given in eq. (2.23) is useful in computing the
amplitudes of the Wilson loops. However, the formula (2.23) is just formal,
since we need still to specify also the boundary conditions of $G(z,w)$.
We remember
in fact that the Lorentz gauge fixing, which in complex coordinates has
the form
$\partial_zA_{\bar z}+\partial_{\bar z}A_z=0$,
allows a residual gauge invariance given by the transformations of the kind:
$$\cases{A^T_{\bar z}\rightarrow A_{\bar z}^T+\partial_{\bar z}\rho\cr
A_z^T\rightarrow A_z^T+\partial_z\rho\cr}\eqno(2.24)$$
where $\partial_z\partial_{\bar z}\rho=0$.
On a compact Riemann surface the only possible solution to this equation is
$\rho={\rm const}$.
The residual gauge invariance amounts in this case to the shift $\varphi
\rightarrow \varphi+{\rm const}$. As in the usual bosonic string theory,
this is not a serious problem and can be solved choosing a normalization of
the propagator $G(z,w)$ at some point. Alternatively one can compute just
the physical amplitudes which should be invariant under translations of the
scalar fields.
One example is provided by the Wilson lines:
$${\rm exp}\left[ic\int_a^bA_\mu dx^\mu\right]=
{\rm exp}\left[ic(\varphi(b)-\varphi(a))\right]\times{\rm c.c.}$$
In order to get rid of the residual gauge invariance on
a noncompact manifold like the complex plane, one has to require boundary
conditions like
$A_z=A_{\bar z}=0$ when $z\rightarrow \infty$.
This means that
$$\partial_zG(z,w)=\partial_{\bar z}G(z,w)=0\eqno(2.25)$$
at $z=\infty$.\smallskip
If the manifold has a boundary $\Omega$, the boundary conditions
should be chosen in such a way that no harmonic functions satisfying
$\triangle_z\rho=0$ are possible. On a disk, for example, this purpose
is achieved requiring eq. (2.25) to be valid on the boundary.
Finally, additional problems can arise in constructing the biharmonic
correlation functions $G(z,w)$ on a noncompact manifold $M$ [7].
In certain cases the direct construction of the propagator $<A_z^TA_w^T>$
in the following way is preferable:
$$<A^T_zA^T_w>=\int_Md^2tg_{t\bar t}\left(R_z(z,t)R_w(w,t)\right)\eqno(2.26)$$
where $\partial_{\bar z}R_z(z,t)=\delta^{(2)}_{\bar z z}(z,t)$.
\vfill\eject
\centerline{\title 3. THE MAXWELL FIELD THEORY ON A FLAT SPACE-TIME}
\vskip 1cm
We begin to study the theory of eq. (2.17) of complex scalar fields
with higher derivatives. This theory is equivalent to the MFT after the
Lorentz gauge fixing has been chosen.
First of all we treat the simple case in which the
topology of the 2D space-time is just that of the complex plane $\rm\bf C$.
In order to solve the free equations of motion (2.21), i.e. with $J_\mu=0$,
we consider the auxiliary scalar field
$${\tilde \varphi}(z,{\bar z})=
g^{z\bar z}\triangle_z\varphi(z,{\bar z})\eqno(3.1)$$
${\tilde \varphi}(z,{\bar z})$ satisfies an harmonic equation:
$$\triangle_z{\tilde \varphi}(z,{\bar z})=0$$
whose solutions are:
$${\tilde\varphi}(z,{\bar z})=i\left (u(z)+{\bar u}({\bar z})\right )$$
$u(z)=\sum\limits_{n=0}^\infty a_n z^n$ being an harmonic function.
The factor $i$ in the above equation is necessary since ${\tilde\varphi}
(z,{\bar z})$ is defined as a purely imaginary field from eq. (2.13).
At this point we can invert eq. (3.1) and obtain the final solution of eq.
(2.21):
$$\varphi(z,{\bar z})=\int_p^z\int_{\bar p}^{\bar z}dt d{\bar t}
g_{t\bar t}\enskip i\left ( u(t)+{\bar u}({\bar t})\right )+i\xi(z,{\bar z})
\eqno(3.2)$$
where $\xi(z,{\bar z})$ is a real harmonic function and $p$ represents an
arbitrary basepoint.
If $g_{t\bar t}=1$, then eq. (3.2) gives for $p=0$:
$$\varphi(z,{\bar z})=iz{\bar z}
(h(z)+{\bar h}({\bar z}))+i\xi(z,{\bar z})$$
This is the most general solution of the biharmonic equation in the flat case.
Eq. (3.2) represents all the zero modes of the operator $\triangle_z
g^{z\bar z}\triangle_z$ on the complex plane with an arbitrary metric.
Let us notice that, due to the presence of the metric, the equations of motion
(2.22) are invariant under conformal transformations in the following sense.
After a conformal transformation $z=z(w)$,
we have $\triangle_zg^{z\bar z}\triangle_z=\left |{dw\over dz}\right|^2
\triangle_wg^{w\bar w}\triangle_w$ which is the same operator
as before.
However the new metric is now
$g^{w\bar w}(w,{\bar w})=\left|dw/dz\right|^2g^{z\bar z}(z,{\bar z})$.
Therefore eqs. (2.17)-(2.18) do not describe a conformal field theory.
This is in agreement with the fact that MFT is conformal only in four
dimensions.
One of the consequences is that the energy momentum tensor is not traceless.
To show this, we make the substitutions (2.14) in the general expression
of the energy momentum tensor of MFT on a curved background:
$$T_{\mu\nu}=-F_{\mu\alpha}F^\alpha_\nu-{1\over 4}g_{\mu\nu}F_{\rho\sigma}
F^{\rho\sigma}\eqno (3.3)$$
The result is:
$$T_{zz}=T_{{\bar z}{\bar z}}=0\qquad\qquad T_{z{\bar z}}=g^{z\bar z}
\triangle_z\varphi\triangle_z\varphi$$
Therefore the energy momentum tensor has a nonvanishing trace. Moreover it
consists in a pure trace term.\smallskip
Let us now compute the propagator $G(z,w)$ from eq. (2.22).
We remember that, on the complex plane, the $\delta$ function has the following
expression:
$$\delta_{z\bar z}^{(2)}(z,w)=\triangle_z{\rm log}\left |z-w\right|^2
\eqno(3.4)$$
Substituting eq. (3.4) in eqs. (2.22) we get:
$$\cases{\triangle_zG(z,w)=g_{z\bar z}{\rm log}\left|z-w\right|^2\cr
\triangle_wG(z,w)=g_{w\bar w}{\rm log}\left|z-w\right|^2\cr}\eqno(3.5)$$
First of all we derive the Green function $G(z,w)$ on a disk
$B_\rho $ of radius $\rho$.
The strategy is to obtain the propagator on the complex plane in the limit
$\rho\rightarrow\infty$.
At the boundary, $|z|=\rho$, there are many possible boundary conditions
for $G(z,w)$ [7], for example $\varphi=\triangle_z\varphi=0$ or
$\varphi=\partial_n\varphi=0$, where $\partial_n$ denotes the
inner normal derivative.
However, in our case, we have to demand that the gauge fixing
condition is true also at the boundary. This implies
$$\triangle_zG(z,w)=\triangle_wG(z,w)=0$$
at the boundary. Moreover it is easy to show that the residual gauge
invariance (2.24) requires the stronger boundary conditions (2.25).
The propagator of the scalar fields with only a singularity in $z=w$ and
fulfilling the boundary conditions (2.25) is the following:
$$G(z,w)=\left|z-w\right|^2{\rm log}\left|
{\rho(z-w)\over \rho^2-\bar w z}\right|^2
+{1\over 2\rho}\left(|z|^2-\rho^2\right)\left(|w|^2-\rho^2\right)\eqno(3.6)$$
Therefore, using eq. (2.23), it is possible to obtain the correct propagator
of the transverse vector fields of the MFT in the Lorentz gauge.
Unfortunately eq. (3.6) is valid only in the case in which the metric
$g_{z\bar z}=1$.\smallskip
Let us now consider the complex plane. Naively one would perform
the limit $\rho\rightarrow\infty$ in eq. (3.6), but it is easy to check
that this limit does not
exist. The difficulties we encounter are part of a more general problem
in defining
the propagator $G(z,w)$ with given boundary conditions on a noncompact
manifold [7]. Here we try to find the propagator of the
vector fields.
{}From eq. (2.20) we need to solve the equation:
$$\partial_{\bar z} g^{z\bar z}\partial_{\bar z}<A_z(z,\bar z)
A_w(w,\bar w)>=-\delta^{(2)}_{z\bar z}(z,w)\eqno(3.7)$$
and simultaneously an analogous equation in $w$.
Due to the fact that on the complex plane the $\delta$ function is
given by $\delta^{(2)}_{z\bar z}(z,w)=\triangle_z{\rm log}\left|z-w\right|^2$,
the solution to eq. (3.7) is:
$$<A_z(z,\bar z)A_w(w,\bar w)>=\int_{\rm\bf C}d^2t\partial_w
{\rm log}\left|t-w\right|^2\partial_z{\rm log}\left|t-z\right|^2
\eqno(3.8)$$
We notice that the propagator (3.8) satisfies the correct boundary
conditions
since it goes to zero in the limit in which $z$ or $w$ go to infinity.
On the contrary, the naive solution $\partial_z\partial_wG(z,w)=
{\rm log}|z-w|^2$, with $G(z,w)=|z-w|^2{\rm log}|z-w|^2-|z-w|^2$, is not
correct since it has the wrong boundary conditions in the above limit.
However, since $\partial_w{\rm log}|t-w|^2\sim1/(t-w)$,
the integral in eq. (3.8) is not well defined.
In fact
$$\partial_w {\rm log}\left|t-w\right|^2
\partial_z{\rm log}\left|t-z\right|^2\sim{1\over t^2}$$
when $t\rightarrow\infty$.
To solve this problem we add to eq. (3.8) an infinite constant which improves
the convergence of the integral.
The following Green function:
$$<A_z(z,\bar z)A_w(w,\bar w)>=\int_{\rm\bf C}d^2t\partial_w
{\rm log}\left|t-w\right|^2\partial_z{\rm log}\left|t-z\right|^2-2
\int_a^{+\infty}{dx_2\over x_2}-2\int_a^{+\infty}{dx_1\over x_1}$$
with $a>0$ being a real constant and where we have used the notation of
eq. (2.4), leads to a well defined propagator.
\smallskip
Now we return to the MFT in order to compute the VEV's of the gauge invariant
quantities, i.e. the Wilson loops $W(L)$, where $L$
is a closed path on $\rm\bf C$.
Taking into account the fact that in the Lorentz gauge
it remains just one degree of freedom $A^T_z$, we have:
$$W(L)={\rm exp}\left [-c^2\oint_L\oint_Ldzdw<A^T_z(z,{\bar z})A^T_w(w,{\bar
w}>
\right]\eqno(3.9)$$
Due to the form of the propagator of the transverse fields given in eq.
(2.23), the VEV of the Wilson loop on the disk is trivial, i.e. $W(L)=1$.
In the case of the complex plane the result is surely
trivial if the radial metric in eq. (3.8)
is defined with $\alpha<-2$. However, when $0>\alpha\ge-2$, the integration in
$d^2t$ does not commute with the integration over the Wilson
loops so that it is not possible to perform the calculation
of the VEV explicitly.
\vskip 1cm
\centerline{\title 4. THE MAXWELL FIELD THEORY ON A RIEMANN SURFACE}
\vskip 1cm
As a first step we consider the MFT on a complex sphere ${\rm\bf CP}_1$
of genus $g=0$.
With respect to the complex plane, the sphere ${\rm\bf CP}_1={\rm\bf C}\cup
\{\infty\}$ includes also the point at infinity.
The sphere ${\rm\bf CP}_1$ is covered here
by two open sets $U,U'$ containing the
points $0$ and $\infty$ respectively.
Local coordinates on $U$ and $U'$ are $z$ and $z'$. When the two open sets
overlap, i.e. $U\cap U'\ne \emptyset$, the two systems of coordinates are
related as follows: $z=1/z'$.
Choosing the metric $g_{z\bar z}$ on $U$ we have
$$g_{z\bar z}=g_{z'\bar z'}\left|z\bar z\right |^{-2}=
g_{z'\bar z'}\left|z'\bar z'\right |^2\eqno(4.1)$$
In the sense explained
in Section 2, all the classical equations of motion and the
action (2.18) are
invariant under a change of coordinates like that of eq. (4.1).\smallskip
Let us solve the equations of motion (2.21). We proceed as in the case
of the complex plane.
On a sphere the only possible solution of the equation
$\triangle_z\tilde \varphi(z,\bar z)=0$
is $\tilde \varphi(z,\bar z)=i\tilde \varphi_0$, where $\tilde\varphi_0$
is a real constant.
Therefore, inverting eq. (3.1) we get:
$$\varphi(z,\bar z)=i\int_p^z\int_{\bar p}^{\bar z}
dt d\bar t g_{t\bar t}\tilde \varphi_0$$
If we set $g_{t\bar t}=1$ on $U$,
$\varphi(z,\bar z)$ is a linear combination of the
following independent zero modes, $z\bar z$, $z+\bar z$ and $1$.
At $z=\infty$ and supposing that $g_{t\bar t}=1$, the metric becomes
$g_{t'\bar t'}=(t'\bar t')^{-2}$ and the above zero modes take the form
$\left |z'\right|^{-2}$, $1/\bar z'$, $1/z'$, $1$.
Due to the presence of the metric in the operator
$\triangle_zg^{z\bar z}\triangle_z$ these functions are zero modes despite of
their singularities at $z=\infty$.
It is easy to check that no other zero modes are possible.
Using the same procedure as we did in the cases of the disk and of the
complex plane,
it is also possible
to derive an expression of the Green function $G(z,w)$ on the complex
sphere. The only residual gauge invariance is now
given by the translation of the scalar field $\varphi(z)$ by a constant
since the equation $\triangle_z\rho$ has just the trivial solution.
Moreover the $\delta$-function is expressed
in terms of the prime form
$$E(z,w)={\rm log}\left|{z-w\over \sqrt{dz}\sqrt{dw}}\right|^2\eqno(4.2)$$
in the following way:
$$\delta^{(2)}_{z\bar z}(z,w)=\triangle_z{\rm log}\left|E(z,w)\right|^2
\eqno(4.3)$$
The prime form of eq. (4.2) does not have logarithmic singularities when
$z,w\rightarrow\infty$ separately.
The only possible singularity occurs at $z=w$.
{}From eq. (4.3) we get:
$$G(z,w)=\int_{{\rm\bf CP}_1}d^2tg_{t\bar t}{\rm log}\left|E(t,w)\right|^2
{\rm log}\left|E(t,z)\right|^2
\eqno(4.4)$$
A metric on ${\rm\bf CP}_1$ leading to a well defined integral is for
example
$$g_{z\bar z}dzd\bar z={dzd\bar z\over(1+z\bar z)^2}$$
Again the computation of the Wilson loop in eq. (3.8) leads to the trivial
result $W(L)=1$.
Finally we treat the MFT on a general Riemann surface $\Sigma_g$ of genus $g$.
As in the case of the sphere, the only possible zero modes of eq. (2.21) are
of the form:
$$\varphi(z,\bar z)=i\int_p^z\int_{\bar p}^{\bar z}dtd\bar t g_{t\bar t}
\enskip\tilde \varphi_0\eqno(4.5)$$
\smallskip
In computing the propagator $G(z,w)$ we use the notation of [10].
Let us consider the prime form:
$$E(z,w)={\theta\left[\matrix{\vec a_0\cr \vec b_0\cr}\right]\left(\int_w^z
{\vec\omega}\right)\over h(z)h(w)}\eqno(4.6)$$
where $\vec a_0$, $\vec b_0$ describe the period of an odd spin structure,
$\vec \omega$ is the vector $(\omega_1,\ldots,\omega_g)$ having as components
the holomorphic differentials $\omega_i(z)$ and finally:
$$h(z)=\sum\limits_{i=1}^g\omega_i(z)\partial_{e_i}\theta
\left[\matrix{\vec a_0\cr \vec b_0\cr}\right](e_i)|_{e_i=0}$$
A representation of the $\delta$-function on a Riemann surface is given by:
$$\delta_{z\bar z}^{(2)}(z,w)=\triangle_z\left[{\rm log}\left|
E(z,w)\right|^2+{\pi\over 2}R(z,w)\right]\eqno(4.7)$$
where
$$R(z,w)=\sum\limits_{i,j=1}^g\left(\omega(z)-\bar{\omega(\bar z)}\right)_i
\Im\left[\Omega\right]^{-1}_{ij}\left(\omega(w)-\overline
{\omega(\bar w)}\right)_j
\eqno(4.8)$$
and $\Omega$ is the period matrix defined in Section 2.
Let us notice that the function
$$K(z,w)={\rm log}\left|E(z,w)\right|^2+{\pi\over 2}R(z,w)\eqno(4.9)$$
is invariant under modular transformations and it is singlevalued around
the nontrivial homology cycles.
It is easy to see that the propagator of the scalar fields is provided by:
$$G(z,w)=\int_{\Sigma_g}d^2tg_{t\bar t}K(z,t)K(w,t)\eqno(4.10)$$
First of all, for a sufficiently regular metric $g_{t\bar t}$, the integral
of eq. (4.12) is well defined and does not diverge. In fact
$\Sigma_g$ is compact and can be represented as a
polygon whose sides are the homology cycles. Inside this polygon
the integrand of eq. (4.10) is a square integrable function apart from the
singularities at the points
$t=z$, $t=w$. Hence it fulfills the existence conditions given in [7].
Moreover, remembering that
$$\triangle_zK(z,t)=\delta_{z\bar z}^{(2)}(z,t)-
{g_{z\bar z}\over N}$$
where
$$N=\int_{\Sigma_g}d^2tg_{t\bar t}$$
we have:
$$\triangle_zG(z,w)=g_{z\bar z}K(z,w)-\int_{\Sigma_g}d^2tg_{t\bar t}K(w,t)
g_{z\bar z}$$
Applying the operator $\triangle_zg^{z\bar z}$ on the above tensor
we have the desired result:
$$\triangle_zg^{z\bar z}\triangle_zG(z,w)=\delta_{z\bar z}^{(2)}(z,w)
-{g_{z\bar z}\over N}$$
The term $g_{z\bar z}/N$ appears due to the residual gauge invariance
$\varphi\rightarrow\varphi+{\rm const.}$ as discussed in section 2.
Computing the VEV of a WL on a Riemann surface one finds that the
contributions
of the transverse gauge fields is trivial due to eq. (2.23). It remains the
integration over the harmonic pieces, which yields a vanishing result as
explained in [4].
\vfill\eject
\centerline{\title 5. QUANTUM ELECTRODYNAMICS IN TWO DIMENSIONS}
\vskip 1cm
In the following we consider the two dimensional MFT discussed above coupled
to massless fermions (Schwinger model [11]):
$$S_{{\rm QED}_2}[A,\bar\psi,\psi,J,\xi]=\int_Md^2z\left[{g_{z\bar z}\over2}
F_{z\bar z}F^{z\bar z}+\bar \psi_\theta(\partial_{\bar z}+A_{\bar z})
\psi_\theta+\bar \psi_{\bar \theta}(\partial_z+A_z)\psi_{\bar \theta}+\right.
$$$$\left.J_zA_{\bar z}+J_{\bar z}A_z+g_{\theta\bar \theta}(\xi_\theta
\psi_{\bar \theta}+\xi_{\bar \theta}\psi_\theta)+
g_{\theta\bar\theta}(\bar\xi_\theta\bar\psi_{\bar\theta}+\bar\xi_{\bar\theta}
\bar\psi_\theta)\right]\eqno(5.1)$$
where $\xi$, $\bar\xi$ are the external currents related to the fields $\psi$,
$\bar\psi$ respectively and $g_{\theta\bar\theta}=\sqrt{g_{z\bar z}}$.
The spinor indices are denoted with $\theta$, $\bar\theta$.
We assume that the ``physical" boundary conditions for $\psi$ and $\bar\psi$
when transported
along the homology cycles in the case in which $M=\Sigma_g$ are given by
the even spin structure $m=\left[\matrix{\vec a_0 \cr \vec b_0\cr}\right]$.
$\vec a_0$ and $\vec b_0$ are two vectors of dimension g whose elements are
half integers such that $4\vec a_0\cdot \vec b_0=0$ mod $2$ [12].\smallskip
Despite of the fact that QED is, at least in the flat case,
solvable in two dimensions [13], we will use here a perturbative
approach.
The applications of the perturbation theory to the Schwinger model were
investigate for example in [14].
Our aim is the computation of the Green functions of ${\rm QED}_2$
at least at the lowest order.
Exploiting the Hodge decomposition of eq. (2.14), eq. (5.1) becomes:
$$S[\varphi,\rho,\bar\psi,\psi]=S_0+S_{\rm I}+S_{\rm z.m.}+S_{\rm c}
\eqno(5.2)$$
where $S_0$ is the free field action:
$$S_0[\varphi,\bar\psi,\psi]=
\int_Md^2z\left[g^{z\bar z}
\triangle_z\varphi\triangle_z\varphi+\bar\psi_\theta\partial_{\bar z}
\psi_\theta+\bar\psi_{\bar\theta}\partial_z\psi_{\bar\theta}\right]
\eqno(5.3a)$$
and $S_{\rm I}$ contains the interaction part:
$$S_{\rm I}[\varphi,\rho,\bar \psi,\psi]=-\int_Md^2z\varphi
\left[\partial_z(\bar\psi_{\bar \theta}\psi_{\bar\theta})-\partial_{\bar z}
(\bar \psi_\theta\psi_\theta)\right]-
\int_Md^2z\rho\left[\partial_z(\bar\psi_{\bar \theta}
\psi_{\bar\theta}
)+\partial_{\bar z}(\bar\psi_\theta\psi_\theta)\right]\eqno(5.3b)$$
Moreover
$$S_{\rm z.m.}=\int_Md^2zA_z^{\rm har}
\bar\psi_{\bar\theta}\psi_{\bar \theta}+
\int_Md^2zA_{\bar z}^{\rm har}\bar\psi_\theta
\psi_\theta\eqno(5.3c)$$
$j_z=\bar\psi_\theta\psi_\theta$ and $j_{\bar z}=\bar \psi_{\bar\theta}
\psi_{\bar\theta}$ are the currents associated with the conservation of
the number of fermions.
We can use the Hodge decomposition (2.15) also for $j_z$, $j_{\bar z}$ and,
exploiting the orthogonality property of this decomposition, it is easy to
see that $S_{\rm z.m.}=\int_M A_z^{\rm har}j_{\bar z}^{\rm har}+{\rm c.c.}$
Therefore $S_{\rm z.m.}$ does not contribute to the computation of the
Green functions, but becomes relevant in the computation of the partition
function [15].
Finally $S_{\rm c}$ contains the external currents.
Again, due to the properties of the Hodge decompositions (2.14)-(2.15),
the zero modes
do not yield a relevant contribution in this sector so that we can write:
$$S_{\rm c}[J,\xi,\bar\xi]=\int_Md^2z\left[(A_z^TJ_{\bar z}^T+
A_z^LJ_{\bar z}^L+{\rm c.c.})+g_{\theta\bar\theta}(\xi_\theta\psi_{\bar\theta}
+\bar\xi_\theta\bar\psi_{\bar\theta}+{\rm c.c.})\right]\eqno(5.3d)$$
The generating functional of the ${\rm QED}_2$ Green functions becomes:
$$Z[J,\bar\xi,\xi]=\int D\varphi D\rho D\bar\psi D\psi\prod\limits_{i=1}^g
d^2a_ie^{-S[\varphi,\rho,\bar\psi,\psi,J,\bar\psi,\psi]}\eqno(5.4)$$
with the action provided by eq. (5.2).
Looking at equations (5.3b) and (5.3d) we see that the integration over
$D\rho$ is not difficult and yields the following result in the limit
in which the external currents $J_z^L$, $J_{\bar z}^L$
are zero:
$$Z[J^T,\bar\xi,\xi]=\int D\varphi D\bar\psi D\psi
e^{-(S_0+S_{\rm i}+S_{\rm c})(\varphi,\bar\psi,\psi,J,\bar\xi,\xi)}
\delta\left(\Re\left[\partial_z(\bar\psi_{\bar\theta}\psi_{\bar\theta})
\right]\right)\eqno(5.5)$$
The $\delta$-function in eq. (5.5) express the physical requirement
that the total number of fermions should be conserved in the amplitudes.
As a matter of fact the equation:
$$\Re[\partial_z(\bar\psi_{\bar\theta}\psi_{\bar\theta})]=\partial_zj_{\bar z}
+\partial_{\bar z}j_z=0\eqno(5.6)$$
means that the current $j_\mu$, $\mu=z,\bar z$, is conserved.
Remembering that in the proper regularization scheme we have
$$<0|\partial_\mu j^\mu f(\bar \psi,\psi,A)|0>=0$$
where $f(\bar \psi,\psi,A)$ is a polynomial in the fields $\bar\psi$,
$\psi$, $A_\mu$, it is clear that we can factor out the $\delta$-function
from eq. (5.5). The factor is an infinite constant $\delta(0)$ which
corresponds to the infinite contribution of the longitudinal degrees of
freedom.
Therefore this way of quantizing the Schwinger model is unusual but consistent.
The alternative is to treat also the scalar fields $\rho$ perturbatively.
In the next section we will present other models in which this is not
necessary.
For example in the action of eq. (6.2), expressing the MFT interacting with
massless scalar fields, the exact forms $\partial_z\rho$ and $\partial_{\bar z}
\rho$ coming from the Hodge decomposition do not even appear.\smallskip
At this point we are ready to compute the amplitudes of ${\rm QED}_2$
perturbatively.
If $M=B_\rho,{\rm\bf CP}_1,\Sigma_g$, the propagators of the scalar fields
are given by eqs. (3.6), (4.4) and (4.10) respectively. The case of the disk
needs some care due to the presence of the boundary.
If $M={\rm\bf C}$, eq. (3.8) provides the propagator of the transverse vector
fields.
Moreover, on $\Sigma_g$, the propagator of the free fermionic fields
with even spin structures is the well known Szeg\"o kernel:
$$S_{\theta\theta'}(z,w)\equiv<\bar\psi_\theta(z)\psi_{\theta'}(w)>=
{\theta\left[\matrix{\vec a_0\cr\vec b_0\cr}\right](z-w)\over
\theta\left[\matrix{\vec a_0\cr\vec b_0\cr}\right](0)E(z,w)}\eqno(5.8)$$
The condition given by eq. (5.6) is trivially satisfied at the lowest order
noting that with the usual regularization [16]:
$$<j_z(z)>=\lim_{z\to w}\left (<\bar\psi(z)\psi(w)>-{1\over z-w}\right)$$
$<j_z(z)>$ is independent of $\bar z$ and has no poles in $z$.
We are now ready to compute the three point function for the coupling
$\varphi\bar\psi\psi$. Here we limit ourselves to the vertex
$V_{z_1\theta_2\theta_3}(z_1,
z_2,z_3)$, which is defined as follows:
$$V_{z_1\theta_2\theta_3}(z_1,
z_2,z_3)=<\varphi(z_1,\bar z_1)\bar\psi_{\theta_2}(z_2,\bar z_2)
\psi_{\theta_3}(z_3,\bar z_3)>\eqno(5.9)$$
The complex conjugate vertex can be computed in an
analogous way.
As in the flat case the three point function of eq. (5.9) is defined by:
$$V_{z_1\theta_2\theta_3}(z_1,
z_2,z_3)=-2{\delta\over\delta J(z_1,\bar z_1)}
{\delta\over \delta\bar\xi_{\bar\theta_2
}(z_2,\bar z_2)}
{\delta\over \delta\xi_{\bar\theta_3}(z_3,\bar z_3)}\times$$$$
\int D\varphi D\bar\psi D\psi
\left.\int_Md^2z'\varphi(z',\bar z')\bar\psi_{\theta'}(z',\bar z')
\psi_{\theta'}(z',
\bar z'){\rm exp}\left[-S_0+S_{\rm c}\right]\right\arrowvert_{\scriptstyle
J=0\atop \bar\xi,\xi=0}\eqno(5.10)$$
where $J$ is given by eq. (2.15).
{}From eq. (5.10) we get the final expression of the vertex
$V_{z_1\theta_2\theta_3}(z_1,
z_2,z_3)$ which is:
$$V_{z_1\theta_2\theta_3}(z_1,
z_2,z_3)=\int_Md^2z'\partial_{\bar z'}G(z',z_1)S_{\theta'\theta_3}
(z',z_3)S_{\theta_2\theta'}(z_2,z')\eqno(5.11)$$
As an upshot the vertex containing the vector field $A_z^T$ becomes:
$$<A_{z_1}^T(z_1)\bar\psi_{\theta_2}(z_2)\psi_{\theta_3}(z_3)>=
\int_Md^2z'\partial_{z_1}\partial_{\bar z'}G(z,z_1)S_{\theta'\theta_3}(z',z_3)
S_{\theta_2\theta'}(z_2,z')\eqno(5.12)$$
Substituting the propagators (4.10) and (5.8) in eqs. (5.11)-(5.12),
we get an expression of the three point function
$V_{z_1\theta_2\theta_3}(z_1,z_2,z_3)$ on a Riemann surface.
On the sphere one can use the propagator (4.4) for the scalar fields
$\varphi$ and the propagator $S_{\theta\theta'}(z,w)=1/(z-w)$ for the
fermions.
In an analogous way one can treat the case of the complex plane.
\vfill\eject
\centerline{\title 6. CONCLUSIONS}
\vskip 1cm
In this paper we have treated the quantization of the two dimensional
Maxwell field theory in the Lorentz gauge and in the presence of a nontrivial
background.
The gauge fixed MFT becomes a theory of scalar fields with higher order
derivatives at least at the tree level. This equivalence is useful since
in this way we can exploit previous mathematical knowledge [7] in order
to construct the propagators of the MFT on two dimensional manifolds.
The propagator of these scalar fields satisfies in fact a biharmonic equation
which has been studied long time ago in connection with a clamped thin plate
subjected to a point load. On a general two dimensional manifold the problem
looses its physical significance but it is still relevant for the
biharmonic classification theory of Riemannian manifolds [7].
To this purpose we notice that the choice of a conformally flat metric in
eq. (2.7) is very useful in treating the MFT in the interacting case.
However, the derivation of the propagator $G(z,w)$ can be easily obtained
also in the case of a general metric. From eq. (2.6) it is in fact clear that
eqs. (3.8), (4.4) and (4.10) are still valid substituting $g_{t\bar t}$
with $1/\sqrt{g}$.
Particular attention has been devoted here to the boundary conditions that the
biharmonic Green functions should satisfy in such a way that the original MFT
remains invariant under the residual gauge transformations.\smallskip
Moreover in the Lorentz gauge the functional integration over the longitudinal
fields is trivial and, due to eq. (2.23), the transverse fields do not give
any contribution to the VEV of the Wilson loops.
Therefore the results of [4,5] are confirmed without using the equations of
motion (2.3) explicitly.\smallskip
Another advantage of having fixed the Lorentz gauge with the method explained
in Section 2 is that one can easily derive the $n$-point functions of the MFT
coupled with other field theories.
In Section 5 we have briefly shown how this is possible in the case of the
Schwinger model.
Unfortunately interesting topics about the Schwinger model like the treatment
of the anomalies [1], the derivation of the partition function [15] and the
integrability on a Riemann surface [13],
have been ignored being outside of the purposes of this paper.\smallskip
In principle one can compute the $n$-point functions of the interacting
MFT perturbatively also for theories which are different from the Schwinger
model.
In particular it is interesting the possibility that the gauge fields $A_z$
and $A_{\bar z}$ can interact with the fields appearing in string theory.
The aim is to construct new string theories and to cancel, at least partially,
the Lorentz and Weyl anomalies of two dimensional (chiral)
conformal field theories as discussed in refs. [17].\smallskip
The simplicity of the Schwinger model in the Lorentz gauge consists in
the fact that the longitudinal fields can be integrated away without requiring
a perturbative treatment.
Other theories in which this is possible as well are the MFT coupled with
the $b-c$ systems of string theory:
$$S[A,b,c]=\int d^2z\left ({1\over 4} g_{z\bar z}F_{z\bar z}F^{z\bar z}+
b_{zz}(\partial_{\bar z}+A_{\bar z})c^z+\bar b_{\bar z\bar z}
(\partial_z+A_z)c^{\bar z}\right)
\eqno(6.1)$$
and the MFT coupled to the scalar fields $X$ of string theory in the
following way:
$$S[A,X]=\int d^2z\left ({1\over 4}g_{z\bar z}F_{z\bar z}F^{z\bar z}+
XF_{z\bar z}+{1\over 2}\partial_zX\partial_{\bar z}X\right)\eqno(6.2)$$
In eq. (6.1) the conservation of the ghost number is anomalous but the anomaly
is a total derivative. Therefore, making
the shift $J'_z\rightarrow J_z+{3\over 8}\partial_z{\rm log}(g_{z\bar z})$
in the
external currents we get the condition that the ghost number is conserved
in analogy with eq. (5.6).
Not so easy is the situation in which the gauge fields are minimally
coupled to the scalar fields as in [18]. In this case the integration
over the longitudinal gauge fields becomes complicated and one has to treat
also the longitudinal degrees of freedom perturbatively.
Finally we already noticed that on a manifold $M$ the usual biharmonic
equation
$$\triangle_z^2\varphi^\mu(z,\bar z)=0\eqno(6.3)$$
is not invariant under general transformations of coordinates and one has to
introduce the operator $\triangle_z g^{z\bar z}\triangle_z$.
In eq. (6.3) we have slightly generalized the previous discussion
allowing the fields $\varphi$ to carry an index $\mu=1,\ldots,N$.
An exception, which merits a little digression, is provided
by the complex sphere because it can be covered by two
open sets with transition functions $z=1/z'$.
As a matter of fact eq. (6.3) is invariant under the transformations of
coordinates
$$z\rightarrow{az'+b\over cz'+d}\qquad\qquad\qquad ad-bc=1\eqno(6.4)$$
generating the group ${\rm SL}(2,{\rm\bf C})$
if $\varphi$ transforms as follows:
$$\varphi(z,\bar z)= J\varphi'(z',\bar z')\eqno(6.5)$$
where $J=\left|{dz\over dz'}\right |$.
Eqs. (6.4)-(6.5) form the socalled Mitchell transformations [19].
As an upshot the following action, dependent on a dimensional parameter
$\gamma$:
$$S=\int d^2z\left[\varphi^\mu(\partial_z\partial_{\bar z})^2\varphi_\mu+
{\gamma\over \varphi^\mu\varphi_\mu}\right]$$
is an example of a gauge invariant field theory with higher order derivatives
[20].
The gauge invariance is preserved also if $\gamma$ is set to zero.
In that limit eq. (6.6) provides one of the simplest gauge theories
with nonabelian group of symmetry.
\vskip 1cm
\centerline{\title ACKNOWLEDGEMENTS}
\vskip 1cm
I would like to thank J. Wess and his group for their warm hospitality at the
Max Planck Institute and at the LMU of Munich.
I also acknowledge many enlightening discussions with H. Grosse and
H. H\"uffel during the
preliminary stages of this work. Moreover I am grateful to M. Mintchev
for a fruitful conversation and for pointing out ref. [21], where the
noninteracting quantum Yang-Mills theory is treated.
Finally I thank L. Bonora, E. Gava, R. Iengo,
N. Manko\v c-Bor\v stnik, J. Sobczyk and G. Sotkov for
many discussions and for their interest in my work.
This work was supported by the Consiglio Nazionale delle Ricerche,
P. le A. Moro 7, Roma, Italy.
\vfill\eject
\centerline{REFERENCES}
\vskip 1cm
\item{[1]}
S. Coleman, {\it Phys. Rev.} {\bf D11} (1975), 3026;
A. Z. Capri, R. Ferrari, {\it Nuovo Cim.} {\bf 62A} (1981), 273;
{\it Journ. Math. Phys.} {\bf 25} (1983), 141; A. K. Raina, G. Wanders,
{\it Ann. of Phys.} {\bf 132} (1981), 404;
S. Donaldson, {\it J. Diff. Geom.}
{\bf 18} (1983), 269;
R. Jackiw, R. Rajaraman, {\it Phys. Rev. Lett.} {\bf 54} (1985), 1219;
L. Faddev, S. Shatashvili, {\it Phys. Lett.} {\bf 183B} (1987), 311;
E. Witten, {\it Comm. Math. Phys.} {\bf 117} (1988), 353;
G. Morchio, D. Pierotti, F. Strocchi, {\it Ann. Phys.} {\bf 188}
(1988), 217;
M. Porrati, E. T. Tomboulis, {\it Nucl. Phys.} {\bf B315} (1989), 615;
T. P. Killingback, {\it Phys. Lett.} {\bf 223B} (1989) 357.\medskip
\item{[2]} M. F. Atiyah, R. Bott, {\it Phyl. Transl. R. Soc. Lond.}
{\bf A308} (1982), 523.\medskip
\item{[3]} D. S. Fine, {\it Comm. Math. Phys.}{\bf 134} (1990),273;
S. G. Rajeev, {\it Phys. Lett.} {\bf 212B} (1988), 203.\medskip
\item{[4]} M. Blau, G. Thompson, {\it Quantum Maxwell Theory on Arbitrary
Surfaces}, Preprint NIKHEF-H/91-08, MZ-TH/91-16.\medskip
\item{[5]} J. Soda, {\it Phys. Lett.} {\bf 267B} (1991), 214.\medskip
\item{[6]} S. Pokorski, {\it Gauge Field Theories}, Cambridge University
Press, 1987.\medskip
\item{[7]} L. Sario, M. Nakai, C. Wang, L. O. Chung,
{\it Lect. Notes in Math.} {\bf 605}, Springer Verlag, 1977.\medskip
\item{[8]} L. Alvarez-Gaum\'e, G. Moore, C. Vafa, {\it Comm. Math. Phys.}
{\bf 106} (1986),1.\medskip
\item{[9]} E. Witten, {\it Comm. Math. Phys.} {\bf 121} (1989), 351.
\medskip
\item{[10]} M. Bonini, R. Iengo, {\it Int. Jour. Mod. Phys.}, {\bf A3} (1988),
841.\medskip
\item{[11]} J. Schwinger, {\it Phys. Rev.} {\bf 128} (1962), 2425.\medskip
\item {[12]} J. D. Fay, {\it Lect. Notes in Math.} {\bf 352}, Springer
Verlag, 1973.\medskip
\item{[13]} J. Barcelos-Neto, A. Das, {\it Zeit. Phys. C} {\bf 32}
(1986), 527;
H. Leutwyler, {\it Phys. Lett.} {\bf 153B} (1985), 65;
F. M. Saradzhev, {\it Phys. Lett.} {\bf 278B} (1992), 449.\medskip
\item{[14]} M. S. Chanowitz, {\it Phys. Lett.} {\bf 171B} (1986), 280.
\medskip
\item{[15]} D. Z. Freedman, K. Pilch, {\it Phys. Lett.}{\bf 213B} (1988), 331.
\medskip
\item{[16]} E. Verlinde, H. Verlinde, {\it Nucl. Phys.} {\bf B288} (1987),
357.\medskip
\item{[17]} A. H. Chamseddine, J. Fr\"ohlich, {\it Two Dimensional
Lorentz-Weyl Anomaly and Gravitational Chern-Simons Theory}, Preprint
ETH-TH/91-48, ZU-TH-30/91; A. M. Po\-lya\-kov, {\it Mod. Phys. Lett.} {\bf A2}
(1987), 899; V. G. Knizhnik, A. M. Polyakov, A. A. Zamolodchikov,
{\it Mod. Phys. Lett.} {\bf A3} (1988), 819;
F. David, {\it Mod. Phys. Lett.} {\bf A3} (1988), 1651;
J. Distler, H. Kawai, {\it Phys. Lett.} {\bf 221B} (1989), 509;
A. H. Chamseddine, {\it Phys. Lett.} {\bf 256} (1991), 379.\medskip
\item{[18]} J. M. F. Labastida, A. V. Ramallo, {\it Chiral Bosons Coupled
to Abelian Gauge Fields}, Preprint CERN-TH-5288/89.\medskip
\item{[19]} G. W. Bluman, R.D. Gregory, {\it Mathematika},
{\bf 32} (1985), 118.\medskip
\item{[20]} J. F. Cari\~nena, C. L\'opez, {\it Int. Jour. Mod. Phys.}
{\bf A7} (1992) 2447.\medskip
\item{[21]} E. Witten, {\it Two Dimensional Gauge Theories Revisited},
Preprint IASSNS-HEP-92/15.\medskip
\bye